\documentclass[longauth]{aa} % for the long lists of affiliations 

\usepackage[normalem]{ulem}
\usepackage{graphicx}
\usepackage{color}
\usepackage{url}
\usepackage[export]{adjustbox}
\usepackage[colorlinks=true, allcolors=blue]{hyperref}
\usepackage{lipsum}
\usepackage{multirow}
\usepackage{xcolor}
\usepackage{orcidlink}
\usepackage{xspace}
\usepackage{soul}
\usepackage{tablefootnote}
\usepackage[switch]{linenoaa} 
\usepackage[referable]{threeparttablex}

\usepackage{txfonts}
\usepackage{natbib}
\usepackage{amsmath}    % Advanced maths commands
\usepackage{amssymb}    % Extra maths symbols

\newcommand{\cyg}{\mbox{Cyg\,X-1}\xspace}
\newcommand{\swiftJ}{\mbox{Swift\,J1727.8$-$1613}\xspace}
\newcommand{\gx}{\mbox{GX\,339$-$4}\xspace}

\DeclareMathOperator{\atantwo}{atan2}

\begin{document}

\title{Variability of X-ray polarization of \cyg} 

\titlerunning{Variability of X-ray polarization of \cyg}

\author{Vadim Kravtsov\inst{\ref{in:UTU}}\orcidlink{0000-0002-7502-3173}
\and Anastasiia Bocharova\inst{\ref{in:UTU}}\orcidlink{0009-0008-1244-3606}
\and Alexandra Veledina\inst{\ref{in:UTU},\ref{in:Nordita}}\orcidlink{0000-0002-5767-7253}  
\and Juri Poutanen\inst{\ref{in:UTU}}\orcidlink{0000-0002-0983-0049}
\and Andrew~K.~Hughes\inst{\ref{in:Oxford}}\orcidlink{0000-0003-0764-0687}
\and Michal~Dov\v{c}iak\inst{\ref{in:Praha}}\orcidlink{0000-0003-0079-1239}
\and Elise~Egron\inst{\ref{in:Cagliari}}\orcidlink{0000-0002-1532-4142}
\and Fabio~Muleri\inst{\ref{in:IAPS}}\orcidlink{0000-0003-3331-3794}
\and Jakub~Podgorny\inst{\ref{in:Praha}}\orcidlink{0000-0001-5418-291X} 
\and Jiři~Svoboda\inst{\ref{in:Praha}}\orcidlink{0000-0003-2931-0742}
\and Sofia~V.~Forsblom\inst{\ref{in:UTU}}\orcidlink{0000-0001-9167-2790}
\and Andrei~V.~Berdyugin\inst{\ref{in:UTU}}\orcidlink{0000-0002-9353-5164}  
\and Dmitry Blinov\inst{\ref{in:FORTH},\ref{in:PhysHeraklion}}\orcidlink{0000-0003-0611-5784}  
\and Joe S. Bright\inst{\ref{in:Oxford}}\orcidlink{0000-0002-7735-5796}  
\and Francesco~Carotenuto\inst{\ref{in:OAR}}\orcidlink{0000-0002-0426-3276}
\and David A. Green\inst{\ref{in:UnivCam}}\orcidlink{0000-0003-3189-9998}
\and Adam~Ingram\inst{\ref{in:NewCastle}}\orcidlink{0000-0002-5311-9078}
\and Ioannis~Liodakis\inst{\ref{in:FORTH}}\orcidlink{0000-0001-9200-4006}
\and Nikos Mandarakas\inst{\ref{in:FORTH},\ref{in:PhysHeraklion}}
\and Anagha P. Nitindala\inst{\ref{in:UTU}}\orcidlink{0009-0002-7109-0202}
\and Lauren Rhodes\inst{\ref{in:Oxford}}\orcidlink{0000-0003-2705-4941}
\and Sergei~A.~Trushkin\inst{\ref{in:SAORAS}}\orcidlink{0000-0002-7586-5856}
\and Sergey~S.~Tsygankov\inst{\ref{in:UTU}}\orcidlink{0000-0002-9679-0793}
\and Ma\"{i}mouna Brigitte\inst{\ref{in:Praha},\ref{in:Charles}}\orcidlink{0009-0004-1197-5935}
\and Alessandro Di Marco\inst{\ref{in:IAPS}}\orcidlink{0000-0003-0331-3259}
\and Noemi~Iacolina\inst{\ref{in:ASI}}\orcidlink{0000-0003-4564-3416} 
\and Henric Krawczynski\inst{\ref{in:Wash}}\orcidlink{0000-0002-1084-6507}
\and Fabio La Monaca\inst{\ref{in:IAPS},\ref{in:TorVergata}}\orcidlink{0000-0001-8916-4156}
\and Vladislav Loktev\inst{\ref{in:UTU},\ref{in:Helsinki}}\orcidlink{0000-0001-6894-871X}
\and Guglielmo~Mastroserio\inst{\ref{in:Milano}}\orcidlink{0000-0003-4216-7936}
\and Pierre-Olivier~Petrucci\inst{\ref{in:Grenoble}}\orcidlink{0000-0001-6061-3480}
\and Maura Pilia\inst{\ref{in:Cagliari}}\orcidlink{0000-0001-7397-8091}
\and Francesco~Tombesi\inst{\ref{in:TorVergata},\ref{in:INAF},\ref{in:INFN}}\orcidlink{0000-0002-6562-8654}
\and Andrzej A. Zdziarski\inst{\ref{in:CAMK}}\orcidlink{0000-0002-0333-2452}
}

\authorrunning{V. Kravtsov et al.} 
 
\institute{Department of Physics and Astronomy, FI-20014 University of Turku, Finland \label{in:UTU} \\
\email{vakrau@utu.fi}
\and Nordita, KTH Royal Institute of Technology and Stockholm University, Hannes Alfv\'ens v\"ag 12, SE-10691 Stockholm, Sweden \label{in:Nordita}
\and Astrophysics, Department of Physics, University of Oxford, Denys Wilkinson Building, Keble Road, Oxford OX1 3RH, UK \label{in:Oxford} 
\and Astronomical Institute of the Czech Academy of Sciences, Bo\v{c}n\'{i} II 1401/1, 14100 Praha 4, Czech Republic \label{in:Praha} 
\and INAF Osservatorio Astronomico di Cagliari, Via della Scienza 5, 09047 Selargius (CA), Italy \label{in:Cagliari}
\and INAF Istituto di Astrofisica e Planetologia Spaziali, Via del Fosso del Cavaliere 100, 00133 Roma, Italy  \label{in:IAPS}  
\and Institute of Astrophysics, FORTH, N. Plastira 100, GR-70013  Heraklion, Greece \label{in:FORTH} 
\and Department of Physics, University of Crete, GR-70013 Heraklion, Greece \label{in:PhysHeraklion} 
\and INAF Osservatorio Astronomico di Roma, Via Frascati 33, 00078 Monte Porzio Catone (RM), Italy \label{in:OAR} 
\and Astrophysics Group, Cavendish Laboratory, 19 J. J. Thomson Avenue, Cambridge, CB3 0HE, UK \label{in:UnivCam} 
\and School of Mathematics, Statistics, and Physics, Newcastle University, Newcastle upon Tyne NE1 7RU, UK \label{in:NewCastle}
\and Special Astrophysical Observatory of the Russian Academy of Sciences, Nizhnij Arkhyz  369167, Karachayevo-Cherkessia, Russia \label{in:SAORAS}  
\and Astronomical Institute, Faculty of Mathematics and Physics, Charles University, V Holešovičkách 2, Prague 8, 18000, Czech Republic \label{in:Charles}
\and Agenzia Spaziale Italiana, via della Scienza 5, 09047 Selargius (CA), Italy \label{in:ASI} 
\and Physics Department, McDonnell Center for the Space Sciences, and Center for Quantum Leaps, Washington University in St. Louis, St. Louis, MO 63130, USA \label{in:Wash} 
\and Dipartimento di Fisica, Università degli Studi di Roma `Tor Vergata', Via della Ricerca Scientifica 1, 00133 Rome, Italy \label{in:TorVergata}
\and Department of Physics, P.\,O. Box 64, FI-00014 University of Helsinki, Finland \label{in:Helsinki}
\and Dipartimento di Fisica, Universit\`a Degli Studi di Milano, Via Celoria 16, 20133 Milano, Italy \label{in:Milano}
\and Universit\'{e} Grenoble Alpes, CNRS, IPAG, 38000 Grenoble, France \label{in:Grenoble} 
\and INAF Astronomical Observatory of Rome, Via Frascati 33, 00040 Monte Porzio Catone, Italy \label{in:INAF}
\and INFN Sezione di Roma `Tor Vergata', Via della Ricerca Scientifica 1, 00133 Rome, Italy \label{in:INFN}
\and Nicolaus Copernicus Astronomical Center, Polish Academy of Sciences, Bartycka 18, PL-00-716 Warszawa, Poland \label{in:CAMK}
}

\abstract{
We present the results of a three-year X-ray, optical, and radio polarimetric monitoring campaign of the prototypical black hole X-ray binary \cyg, conducted from 2022 to 2024. 
The X-ray polarization of \cyg was measured 13 times with the Imaging X-ray Polarimetry Explorer (IXPE), covering both hard and soft spectral states. 
The X-ray polarization degree (PD) in the hard state was found to be $\approx$4.0\%, roughly twice as high as in the soft state, where it was around 2.2\%.
In both states, a statistically significant increase in PD with the energy was found. 
Moreover, a linear relation between PD and spectral hardness suggests a gradual and continuous evolution of the polarization properties, rather than an abrupt change of polarization production mechanism between states.
The polarization angle (PA) was independent of the spectral state and showed no trend with the photon energy. 
The X-ray PA is well aligned with the orientation of the radio jet, as well as the optical and radio PAs.
We find significant orbital changes of PA in the hard state, which we attribute to scattering of X-ray emission at the intrabinary structure.
No significant superorbital variability in PD or PA was found at the period $P_{\rm so} = 294$~d. We detect, for the first time in this source, polarization of the radio emission, with the PA aligned with the jet, and a strong increase of the PD at a transition to the soft state.  
We also find no correlation between the X-ray and optical polarization; if any, there is a long-term anti-correlation between the X-ray PD and the radio PD.
}

\keywords{accretion, accretion disks -- polarization --  stars: black holes -- stars: individual: \cyg -- X-rays: binaries}
\maketitle

\section{Introduction}
\label{sec:intro}

Accretion is an efficient mechanism for heating and extracting energy from matter as it falls toward a compact object.
Energy can be released through various channels, producing distinct X-ray emission signatures (spectral states) observed in these systems.
Black hole X-ray binaries (BHXRBs) are known to swing between two major spectral states, hard and soft, distinguished by the energies at which their emission peaks \citep{Zdziarski2004a}.
The soft state is characterized by dominant thermal emission from the classical optically thick and geometrically thin accretion disk \citep{Shakura1973,Novikov1973}.
Transition to the hard spectral state is marked by spectral shift to a power-law shape, peaking at $\sim100$~keV. 
This emission is believed to be produced by multiple Compton upscatterings of soft seed photons, either from the disk or internally produced synchrotron photons, in a hot optically thin medium surrounding the black hole \citep{Sunyaev1980,Gierlinski1997,Poutanen2009,Veledina2013}.
Determining the size of the hot, rarefied plasma responsible for the bulk of X-ray emission, along with its shape and orientation with respect to the disk, remains one of the key open questions in high-energy astrophysics.

Similar X-ray spectra could be produced by various proposed configurations of the hot medium --- such as slab, wedge, cone, or lamppost geometries --- however, the polarization degree (PD), polarization angle (PA), and their spectral dependence differ significantly between these scenarios \citep{Poutanen1996,Poutanen2018,Dovciak2004,KrawczynskiBeheshtipour2022}.
The launch of the Imaging X-ray Polarimetry Explorer \citep[IXPE;][]{Weisskopf2022} in December 2021 has opened a new window to probe the geometry of X-ray-emitting regions in accreting compact objects. 
Cygnus X-1 (\cyg) was the first hard-state BHXRB studied with IXPE \citep{Krawczynski2022}.

\cyg is a bright persistent X-ray binary, harboring the first discovered \citep{Bowyer1965} and possibly one of the most massive Galactic BHs with $M_{\rm BH} =  21.2 \pm 2.1 M_\odot$  \citep{Miller-Jones2021}, although most recent estimates favor a lower mass of $M_{\rm BH} \approx 14~M_\odot$ \citep{Ramachandran2025}. 
It orbits a supergiant O-type donor star in a 5.6~d orbit.  
The donor star in this system nearly fills its Roche lobe and accretion proceeds via wind \citep{GiesBolton1986,Ramachandran2025}.
The accretion geometry, however, is not steady -- this was first noticed as changes in the X-ray spectra of \cyg, which gave rise to the soft/hard classification of the spectral states of X-ray binaries \citep{Tananbaum1972,Zdziarski2004,Done2007}.
Recent estimates suggest the binary inclination is low, $i=27\fdg5\pm0\fdg8$ \citep{Orosz2011,Miller-Jones2021}.
The system inclination has a strong impact on the expected PD, which is highly sensitive to the presence of axial symmetry in the source.
Sources with spherical symmetry, or those viewed face-on with circular symmetry, produce zero PD, whereas edge-on configurations yield a maximal PD as predicted by a specific model.

IXPE observations \citep{Krawczynski2022} revealed an unexpectedly high polarization, PD~$=4.0\pm0.2$\%, with the PA aligned with the direction of the radio jet \citep{Miller-Jones2021}. 
This finding significantly narrowed down the range of viable models, favoring a geometry in which the X-ray emission region is extended perpendicular to the jet, thereby being consistent with the hot flow models, and ruling out several alternatives, such as a vertically extended (jet-base) or lamppost corona.
A similar PD of about 4\%, with the PA again aligned with the jet, was detected in a BH transient Swift~J1727.8$-$1613 during its hard state \citep{Veledina2023,Ingram2024,Podgorny2024}, providing further support to this shape of the X-ray emitting region.   

The detected PD in \cyg is, however, too high for the known system inclination, and is hard to achieve in any scenario.
Additional assumptions have been employed to achieve PD$\sim4$\% in the models: the inclination of the X-ray emitting region might be higher (by $15\degr$--$30\degr$) than the orbital one \citep{Krawczynski2022}, the hot medium may attain a significant outflow velocity \citep{Poutanen2023}, or the PD produced intrinsically in the source may be boosted by the scattering off the accretion disk wind at large radii \citep{Nitindala2025}.

The higher inclination may result either from a steady warp of the accretion flow caused by a misalignment between the BH spin and orbital axis \citep{Bardeen1972} or from a particular phase of precession of the inner flow. 
Precession of the flow or an accretion disk has previously been considered in the source in the context of super-orbital variability of X-ray, optical and radio fluxes, as well as the X-ray hardness and optical polarization signatures \citep{Kemp1983,Priedhorsky1983,Karitskaya2001,Lachowicz2006,Ibragimov2007,Poutanen2008,Zdziarski2011,Kravtsov2022}.
A substantial misalignment between the BH spin and the orbital axis can arise if the BH received a natal kick during its formation \citep{Fragos2010}.
However, \cyg exhibits a small proper motion relative to the Cygnus OB3 stellar association \citep{Rao2020}, placing strong constraints on the natal kick velocity and limiting the maximum possible misalignment angle to $\sim10\degr$ \citep{Miller-Jones2021}.
This upper limit is sufficient to account for the soft-state X-ray polarization signatures of \cyg, PD~$=2.0\pm0.1$\% and an energy-independent PA aligned with the jet, assuming that returning radiation plays a dominant role in this state \citep{Steiner2024}.
These findings disfavor the steady-warp scenario and instead support the case where the inner accretion flow undergoes precession on super-orbital timescales.
If true, this scenario predicts large variations of the PA with the amplitude exceeding $20\degr$ on the timescale of $\sim300$~d, corresponding to the maximal reported super-orbital period.

In this paper, we present the results of thirteen IXPE observations of \cyg conducted between 2022 and 2024, covering its hard, intermediate, and soft spectral states.
These include six observations analyzed here for the first time.
Details of the IXPE observations, supporting multiwavelength data, and the data reduction procedures are provided in Sect.~\ref{sec:data}.
The results of the comprehensive analysis are presented in Sect.~\ref{sec:results}. 
In Sect.~\ref{sec:discussion}, we discuss the implications of our findings in the context of the accretion geometry and polarization production mechanisms.
We summarize our findings and outline future prospects in Sect.~\ref{sec:conclusions}.

\section{Observations and data reduction}
\label{sec:data}

\subsection{IXPE}

\label{sec:ixpe_reductions}

IXPE is the first satellite dedicated to polarimetric X-ray observations that operates in the 2--8 keV band \citep{Weisskopf2022}. 
It carries three X-ray telescopes, each consists of a Mirror Module Assembly and a polarization-sensitive gas-pixel detector unit \citep[DU;][]{Baldini2021, Soffitta2021, DiMarco2022}, enabling imaging X-ray polarimetry of extended sources and a huge increase of sensitivity for point-like sources.
IXPE provides angular resolution of~$\lesssim30\arcsec$ (half-power diameter, averaged over the three detectors). 
The overlap of the fields of view of the three DUs is circular with a diameter of $9\arcmin$; the spectral resolution is better than 20\% at 6 keV.

\begin{table*}
\centering
\caption{IXPE observing log and measured X-ray polarization.}
\label{tab:observation_data}
    \begin{tabular}{clclccccccc}
        \hline
        \hline
        Epoch & Obs. ID & \multicolumn{2}{c}{Date}  & Orbital Phase  &IXPE Hardness & State & PD & PA & Exposure Time\\
        &&&&&4--8 keV/2--4 keV && (\%) & (deg) & (ks) \\
        \hline
        1 & 01002901 &2022&  May 15 & 0.00--1.00 & 0.569 &H& $4.0 \pm 0.2$ & $-21 \pm 2$ & 241.5\\
        2 & 01250101 &&  June 18 & 0.00--0.37 & 0.610  &H& $3.9 \pm 0.3$ & $-26 \pm 3$ & 86.0\\
        \hline
       3 & 02008201 &2023& May 2 & 0.81--0.89 & 0.280  &S& $2.5 \pm 0.4$ & $-17 \pm 5$ & 20.8\\
       4 & 02008301 &&  May 9 & 0.97--0.09 & 0.271  &S& $2.4 \pm 0.3$ & $-23 \pm 4$ & 30.8\\
       5 & 02008401 && May 24 & 0.72--0.83 & 0.151  &S& $2.1 \pm 0.3$ & $-25 \pm 4$ & 24.6\\
       6 & 02008501 &&  June 13 & 0.31--0.43 & 0.210  &S& $1.5 \pm 0.3$ & $-26 \pm 6$ & 28.8\\
       7 & 02008601 &&  June 20 & 0.39--0.54 & 0.240  &S& $2.1 \pm 0.2$ & $-36 \pm 3$ & 34.2\\
       \hline
       8 & 03002201 &2024&  April 12 & 0.43--0.64 & 0.610  &H& $3.9 \pm 0.5$ & $-25 \pm 4$ & 55.8\\
       9 & 03003101 &&  May 6 & 0.79--0.99 & 0.589  &H& $3.1 \pm 0.5$ & $-28 \pm 5$ & 53.8\\
       10 & 03010001 &&  May 26 & 0.29--0.52 & 0.620  &H& $4.6 \pm 0.5$ & $-28 \pm 3$ & 57.5\\
       11 & 03010101 &&  June 14 & 0.80--0.01 & 0.649  &H& $4.6 \pm 0.6$ & $-33 \pm 3$ & 55.7\\
       12 & 03002599$^a$ &&  October 10 & 0.79--0.01 & 0.389  &S& $2.8 \pm 0.3$ & $-18 \pm 3$ & 55.1\\
       13 & 03002599$^a$ &&  December 12 & 0.19--0.42 & 0.280  &S& $2.8 \pm 0.2$ & $-25 \pm 3$ & 56.2\\
       \hline
    \end{tabular}
    \begin{tablenotes}
            \item $^a$Observations 12 and 13 were parts of one observation ID 03002599, which was then manually split into two parts.\\
    \end{tablenotes}
\end{table*}

\begin{figure*}
\centering
\includegraphics[width=0.65\linewidth]{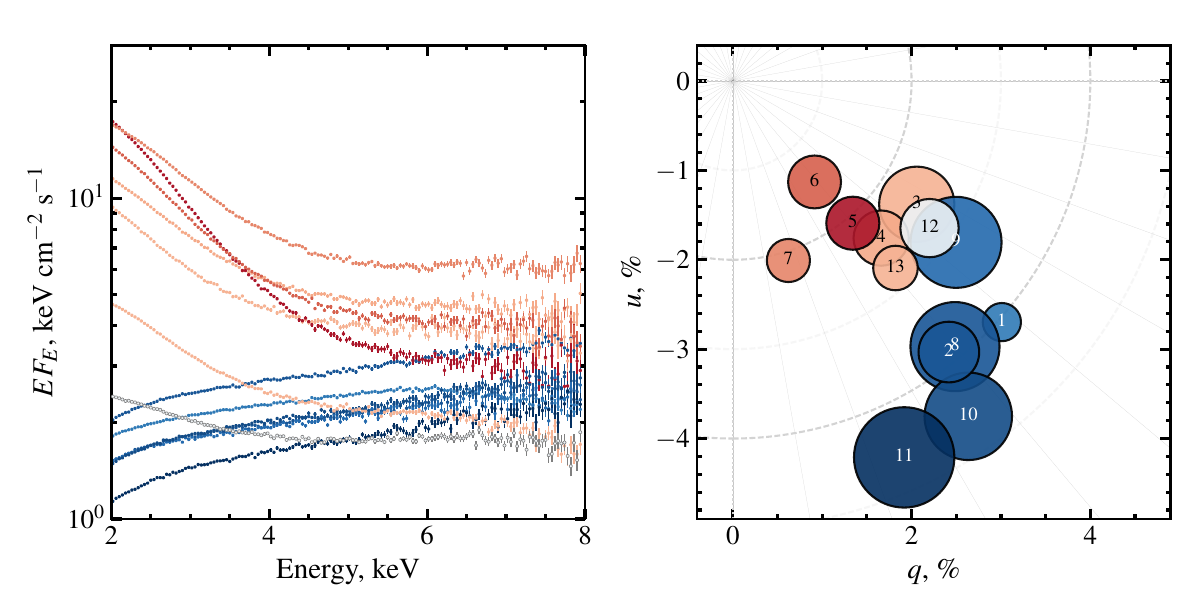}
\includegraphics[width=0.33\linewidth, trim={0.5cm 0 0 0}]{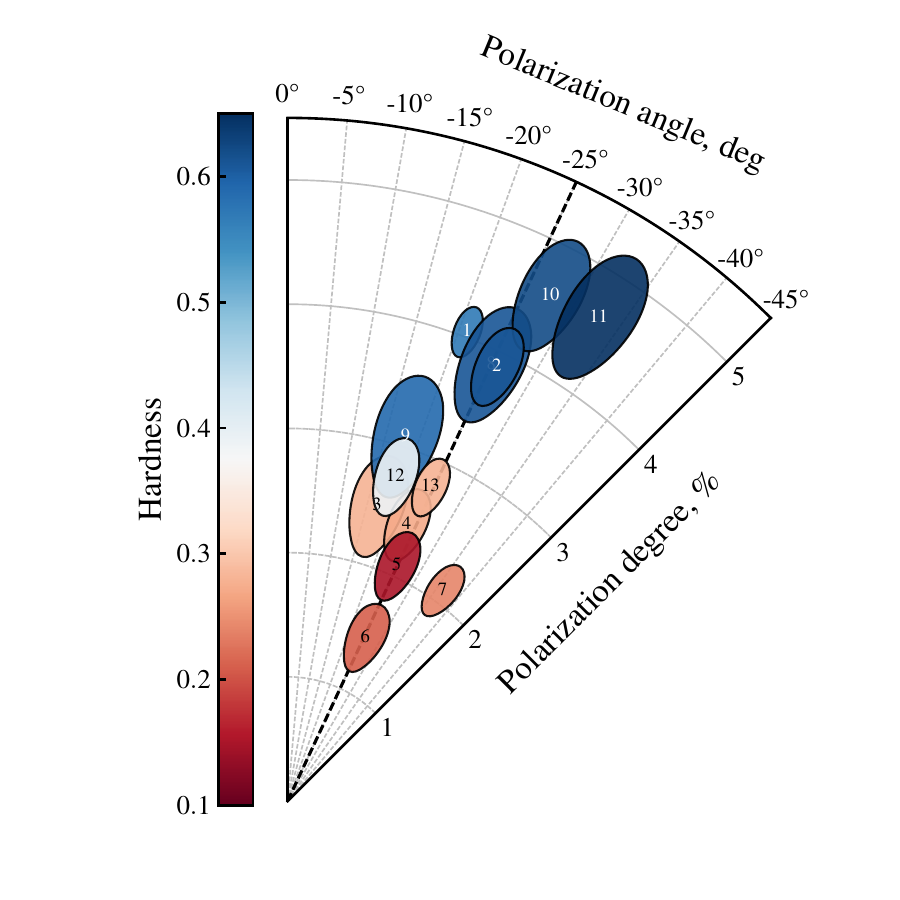}
\caption{Spectral and polarization properties of \cyg for each IXPE observation with hardness shown in color. Numbers represent the observation's epoch in accordance with Table~\ref{tab:observation_data}. IXPE spectra,  normalized Stokes parameters, and PD and PA of X-ray polarization are shown in left, middle, and right panels, respectively. Dashed black line in the right panel indicates the radio jet direction.}
\label{fig:ixpe}
\end{figure*}

We consider the full set of IXPE observations of \cyg, which consists of 13 individual pointings, conducted from 2022 to 2024 (see Table~\ref{tab:observation_data}). 
We used Level 2 data downloaded from the IXPE archive at HEASARC and analyzed it using \textsc{ixpeobssim}  package v31.0.1 \citep{Baldini2022}. 
In the case of the hard state observation in May 2022 we  used post-reconstruction calibration of the energy scale.\footnote{\url{https://heasarc.gsfc.nasa.gov/FTP/ixpe/data/obs/01/01002901/README}}   
The source region for all available observations was defined in \textsc{SAOimageDS9} v8.6 as a circle with radius $60\arcsec$ around the source and extracted using \texttt{xpselect} tool. 
The average polarization properties in the entire energy band for each observation were extracted using the \texttt{xpbin} tool with \texttt{PCUBE} algorithm in a single 2--8 keV energy bin. 
For phase-resolved polarimetric analysis, we utilized the \texttt{xpphase} tool with the orbital period of 5.599829~d and a zero point of JD~2441874.707 \citep{Brocksopp1999b}. We split Epoch 1 data into ten phase bins with \texttt{xpselect} and calculated the polarization properties in each phase bin with the same approach as for the whole observation. 
The hardness ratio was calculated as IXPE photon flux ratio in the energy ranges 4--8 and 2--4 keV. 
The main results are shown in Fig.~\ref{fig:ixpe}. 
Spectra were extracted with the \texttt{xpbin PHA1} algorithm, utilizing CalDB v13 response files.
The superorbital phases were calculated with a period of 294.0~d and a zero point of JD~2440000.0 \citep{Priedhorsky1983}.

\subsection{Multiwavelength coverage}

\begin{figure}
\includegraphics[width=0.85\linewidth, center]{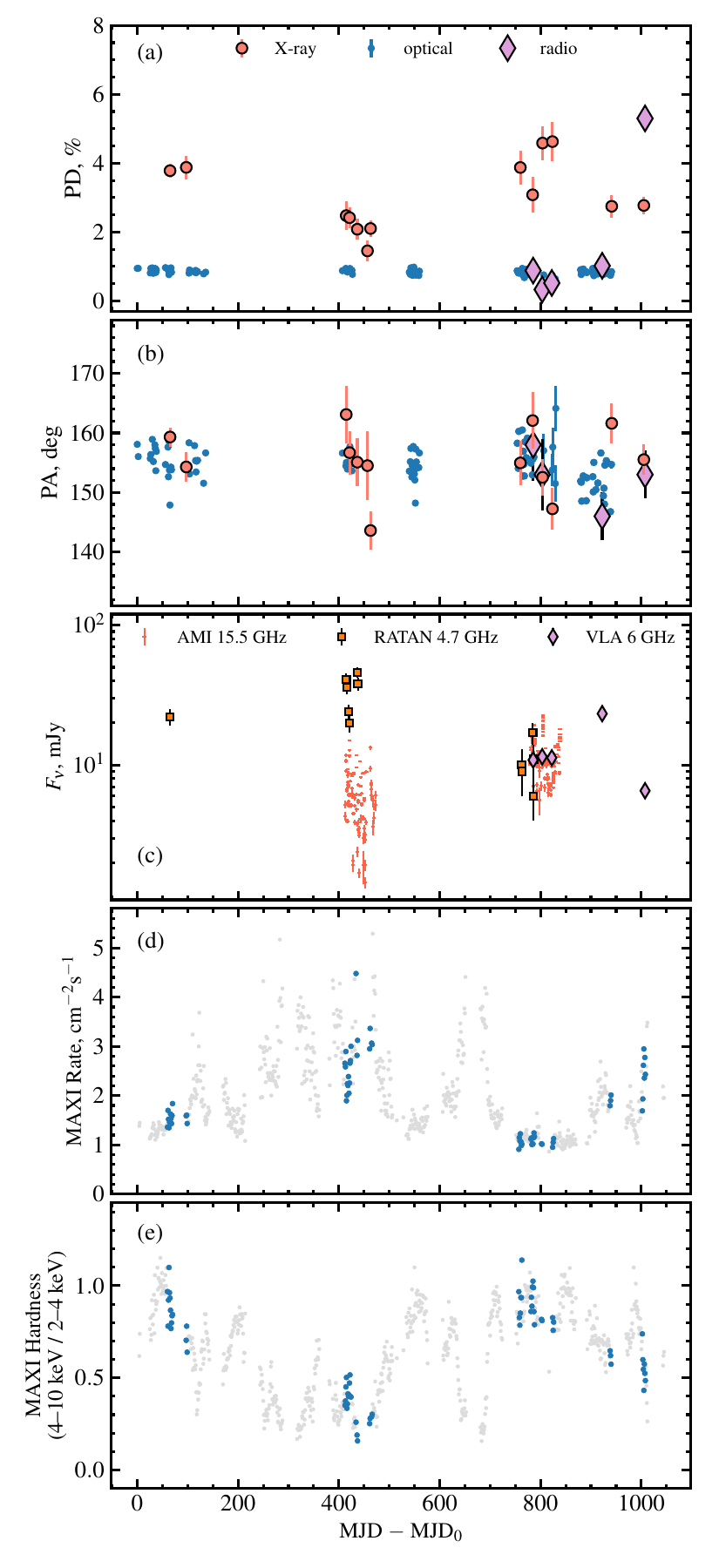}
\caption{Multiwavelength results on \cyg. Panels (a) and (b): PD and PA of polarization in X-rays (IXPE), optical (DIPol) and radio (VLA). Panel (c): light curve in the radio as measured with AMI, VLA and RATAN\,600. Panel (d): MAXI photon flux in 2--20 keV. Panel (e): MAXI hardness. In panels (d) and (e), we show all MAXI data in gray, highlighting in color only the data obtained simultaneously with IXPE. $\rm{MJD}_0 = 59653$ (2022 March 15). }
\label{fig:multiwavelength}
\end{figure}

To support X-ray polarimetric observations with IXPE, the multiwavelength observational campaign was organized with optical and radio facilities. 
The high-precision optical polarimetric observations were performed with DIPol-2/UF instruments at 60 cm Tohoku telescope (T60) at Haleakala Observatory, Hawaii and at 2.56 m Nordic Optical Telescope (NOT), as well as with RoboPol at the Skinakas
Observatory. Radio polarimetric observations were conducted with Karl G. Jansky
Very Large Array (VLA) and where supported with RATAN-600 and the Arcminute Microkelvin Imager (AMI) monitoring.

\subsubsection{Optical observations}

Optical polarimetric observations of \cyg were carried out with the broad-band $BVR$ polarimeters DIPol-2 \citep{Piirola2014} and DIPol-UF \citep{Piirola2021}, mounted on the remotely controlled 60 cm Tohoku telescope (T60) at Haleakala Observatory, Hawaii and at the 2.56 m Nordic Optical Telescope (NOT), Observatorio del Roque de los Muchachos (ORM), La  Palma, Spain, and with RoboPol polarimeter \citep{Robopol} in the focal plane of the 1.3 m telescope of the Skinakas observatory, Greece.
DIPol-2 and DIPol-UF are high-precision double-image CCD polarimeters, capable of measuring polarization simultaneously in three optical ($BVR$) bands. 
The instrumental polarization of both instruments is small ($<10^{-4}$) and is well calibrated by observing 15 to 20 unpolarized standard stars. 
The polarization of the sky is optically eliminated by the design of the instruments. 
The zero point of the PA was determined by observing highly polarized standards HD~204827 and HD~161056.
Each measurement of Stokes parameters took about 20 s and more than 200 individual measurements were obtained during the average observing night.
In total, \cyg was observed for 90 nights in 2022--2024. 
Intrinsic polarization of the source has been extracted by subtracting the Stokes parameters of the interstellar polarization \citep[Table 2 of][]{Kravtsov2023} from the observed Stokes parameters of \cyg.
A more detailed description of the methods and calibrations can be found in \cite{Piirola2020} and \cite{Kravtsov2023}.
The RoboPol data was analyzed by the automatic pipeline described in \cite{Blinov2021} using both polarized and unpolarized standards to characterize the instrumental polarization. 
For the sources not in the central mask, used to account for the ISM polarization,  we used the analysis procedure described in \cite{Panopoulou2015}. 
For consistency, the ISM polarization is corrected using the same reference star for all three instruments.

\subsubsection{The Karl G.~Jansky Very Large Array}

We obtained five radio observations with the Karl G.~Jansky Very Large Array (VLA, Project Code: 24A-469) to track the evolution of the relativistic jets during the IXPE campaign. The observations were conducted on 2024 May 8 (MJD 60438), May 26 (MJD 60456), June 14 (MJD 60475), September 22 (MJD 60575), and December 16 (MJD 60660). The first three epochs were taken in the B configuration, the fourth in a hybrid B$\rightarrow$A configuration, and the final epoch in the most extended A configuration. These configurations provided angular resolutions of $<1\arcsec$, effectively eliminating concerns about source confusion. Each observation used the 3-bit C-band (4–8 GHz) and X-band (8–12 GHz) receivers, yielding approximately 8 GHz of contiguous bandwidth (prior to flagging). The X and C bands were observed consecutively, with X-band preceding C-band in each session. For calibration, we used 3C\,286 as the bandpass, flux scale, and polarization angle (PA) calibrator; J2015+3710 as the complex gain calibrator; and J2355+4950 as the polarization leakage calibrator.
 
We calibrated and flagged the parallel-hand visibilities (i.e., Stokes $I$ and $V$) using the automated VLA pipeline provided in \textsc{casa} v6.5 \citep{Casa2022}. After each pipeline run, we manually inspected the resulting calibrated visibilities and removed any residual corrupted data that had not been caught by the automatic flagging routines. Because the pipeline does not include polarization calibration, we manually calibrated the cross-hand visibilities (i.e., Stokes $Q$ and $U$), following the standard procedures outlined in the VLA calibration guides\footnote{\url{https://casaguides.nrao.edu/index.php/Karl_G._Jansky_VLA_Tutorials}}. Imaging was performed with \textsc{wsclean} \citep{wsclean}, which generated, for each epoch, band, and Stokes parameter, a set of 32 frequency-resolved images (evenly spaced in frequency) as well as a single integrated Multi-Frequency Synthesis (MFS) image to enhance sensitivity. To extract the full polarization flux densities of \cyg, we used the \texttt{imfit} task in \textsc{casa} to model the source in each image as an elliptical Gaussian. Since the source was unresolved, we fixed the Gaussian shape to match the synthesized beam. We estimated the uncertainty in the flux measurements as the root-mean-square (RMS) noise in a nearby, emission-free region covering an area of approximately 100 synthesized beams.
 
Finally, we extracted the key polarization properties -- the PD, PA, and rotation measure (RM) -- using the rotation measure synthesis software contained within the \textsc{rm-tools} \citep{rmtoolscite}; we direct interested readers to \citet{Brentjens2005} for a description of RM synthesis. Initially, we calculated the polarization properties of each band separately, and found no significant frequency-dependency on the measured RM or intrinsic PA$_0$ (i.e., the PA corrected for the effects of Faraday rotation). Moreover, despite the non-simultaneity of our C- and X-band observations, we found that PA$_0$ and RM did not exhibit intra-observation variability. As a result, our reported values of PA$_0$ and RM combine the C- and X-band data to increase signal-to-noise and decrease the error on the inferred RMs.\footnote{\cyg\ was not detected during our fourth observation on 2024 September 22, likely due to secular evolution of the source associated with the transition from C- to X-band. Consequently, we excluded it from the RM synthesis analysis, as its inclusion reduced the significance of the polarization detection.} We applied no further manipulation of our radio observations.

\subsubsection{RATAN\,600}

We carried out monitoring of \cyg with the RATAN\,600 radio telescope at 4.7 GHz and 11.2 GHz from 2022 May to 2024 June using the "Southern Sector and Flat mirror" antenna. 
The sensitivity of such measurements is about 3--10 mJy beam$^{-1}$. 
Thus, \cyg was undetected most of the time, with detections presented in Fig.~\ref{fig:multiwavelength}.
Previous monitoring observations of \cyg have shown typical flux variations in the vicinity of 10--30 mJy at 4.7~GHz. 
Calibration was performed using quasar 3C~48, adopting a brightness of 5.8 and 3.42~Jy at 4.7 and 8.2~GHz, respectively, according to the flux density scale by \cite{Ott1994}. 

\subsubsection{AMI}

\cyg was observed 98 times during May and June 2023 and May
and June 2024 with the AMI Large Array \citep{Zwart2008, Hickish2018} at 15.5 GHz. 
The observations were typically $\sim25$ minutes, with two ten-minute scans of \cyg interleaved between short observations of a nearby compact source. 
The flux density scale of the observations was set by using daily short observations of 3C 286, and the interleaved calibrator observations were used to calibrate antenna-based amplitude and phase variations during the observations. 
The observations covered a 5 GHz bandwidth of a single linear polarization, Stokes $I-Q$.

\section{Results}
\label{sec:results}

\begin{figure}
\centering
\includegraphics[width=0.85\linewidth]{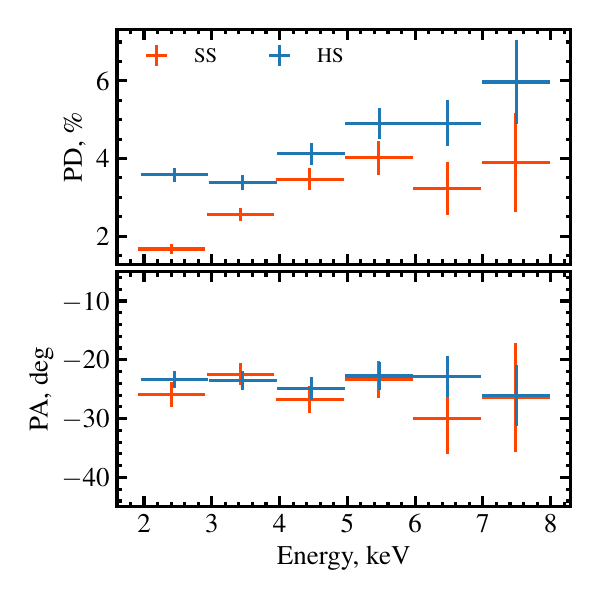}
\caption{Energy dependence of polarization properties of \cyg in the hard  and soft states.}
\label{fig:states_spectra}
\end{figure}

\begin{figure}
\centering
\includegraphics[width=0.85\linewidth]{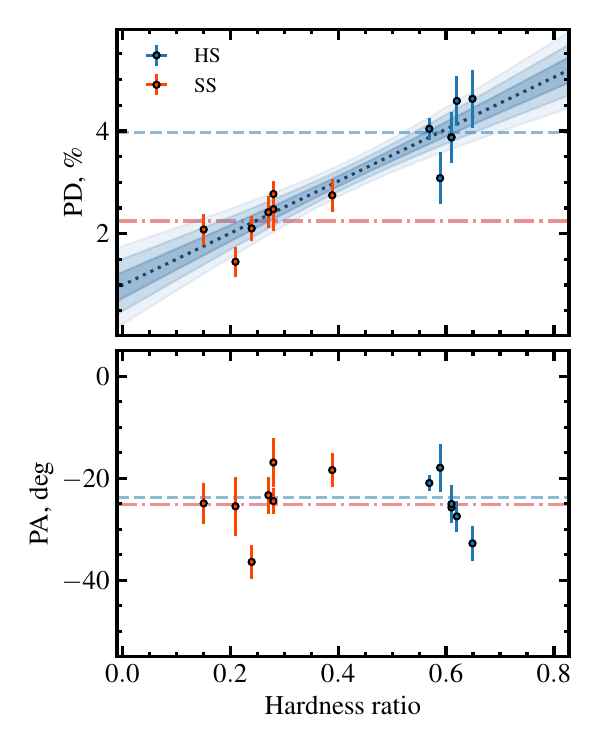}
\caption{Dependence of PD and PA of \cyg on spectral hardness. The dashed and dash-dotted horizontal lines show average values in the hard- and soft-states, respectively. The dotted line with 1, 2, and 3$\sigma$ confidence intervals shows a linear fit to the PD.}
\label{fig:pol_vs_hardness}
\end{figure}

\subsection{Long-term X-ray polarization changes}
\label{sec:long-term_behavior}

We first consider the long-term changes of X-ray polarization by averaging the polarization signatures within each observational epoch.
The {\sc pcube}-average polarization, along with IXPE hardness and assigned states are reported in Table~\ref{tab:observation_data}.
We adopt a hardness value of 0.4 as the threshold between the hard and soft spectral states; this simplified classification does not treat the intermediate state as a separate spectral state.

In Fig.~\ref{fig:ixpe} we show the spectra of individual epochs and their polarization in the $(q,u)$-plane and in the PD--PA plot, color-coded according to spectral hardness.
The PA deviates from the jet position angle \citep[of  $\sim$$-25\degr$;][]{Stirling2001,Miller-Jones2021} by no more than $10\degr$.
The PD is relatively stable within each state, yet clearly depends on the hardness, taking values between $\approx$1.5\% in the soft state to $\approx$4.5\% in the hard state.

The energy dependence of polarization, averaged separately over the hard and soft states, is shown in Fig.~\ref{fig:states_spectra}. 
The PD grows with energy in both states, as indicated by the highly statistically significant improvement of the fit with linearly increasing PD with energy comparing to the constant PD model: the $\chi^2$/d.o.f. value drops from 27.5/5 to 6.3/4  in the hard state and from 78/5 to 7/4 in the soft state.
We further investigate the dependence of PD and PA on hardness in Fig.~\ref{fig:pol_vs_hardness}.
We find that the X-ray PD shows strong positive correlation (Pearson correlation coefficient $r=0.92$) with spectral hardness, while PA is not sensitive to spectral changes -- fit of the linear model to the PA gives $\chi^2$/d.o.f.=13.6/11.

\begin{figure*}
\centering
\includegraphics[width=0.4\linewidth]{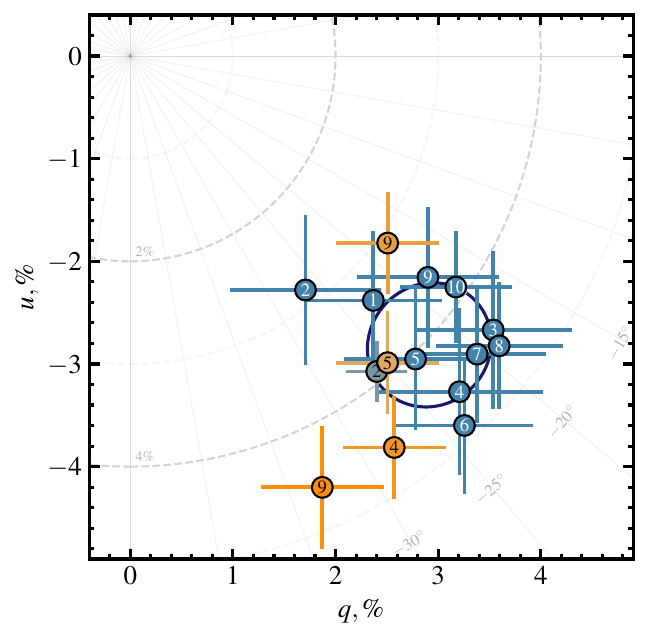}
\includegraphics[width=0.435\linewidth]{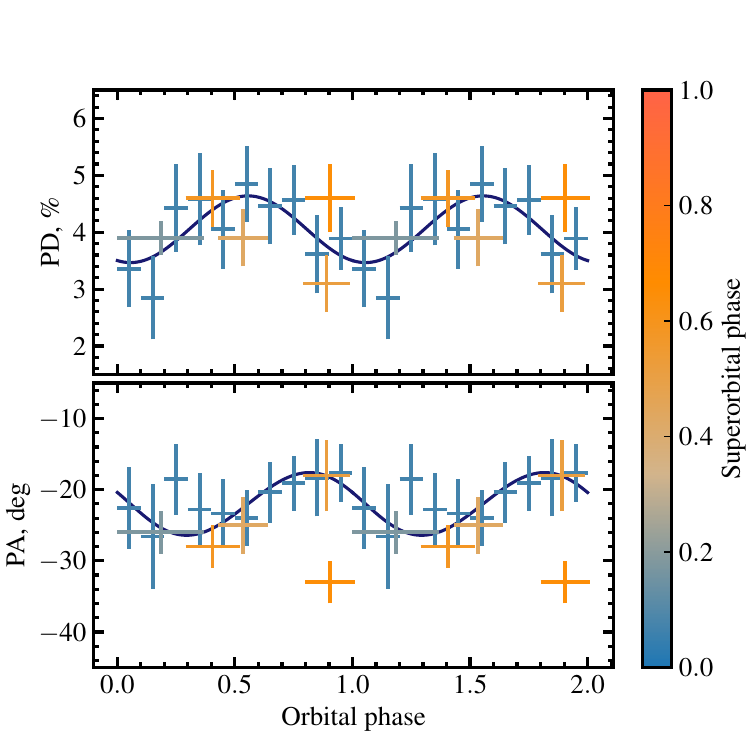}
\caption{Orbital phase-resolved polarization of of \cyg in the hard state. 
Left panel: Variation of the normalized Stokes parameters. The numbers represent the orbital phase bins from 1 to 10  and the colors correspond to the superorbital phase (right color bar). 
Right panel: PD and PA as a function of orbital phase.  
Solid black lines in both panels correspond to the best-fit model described in Sect.~\ref{sec:orb_variations}.
}
\label{fig:phase_resolved_obs}
\end{figure*}

\begin{figure}
\centering
\includegraphics[width=0.9\linewidth]{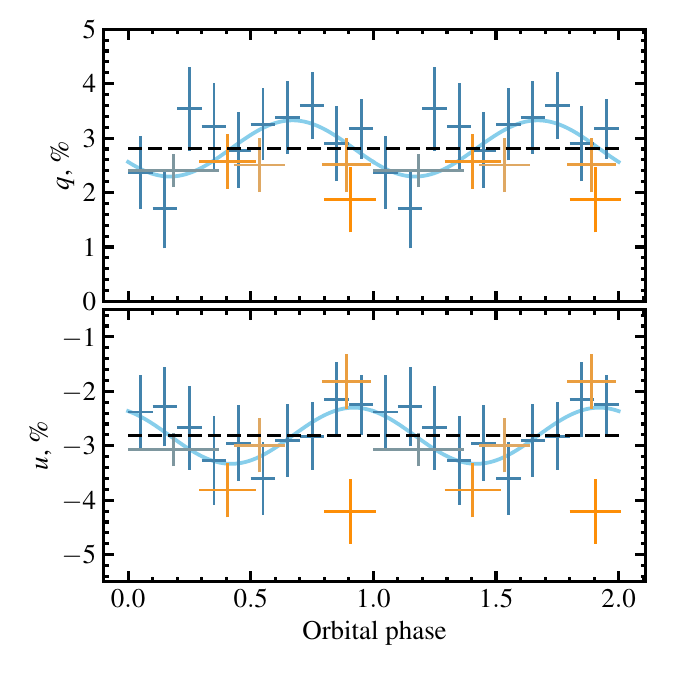}
\caption{Orbital phase-resolved normalized Stokes parameters of \cyg in the hard state. Dashed black and solid blue lines correspond to best-fit constant and sinusoidal models, respectively.
See Sect.~\ref{sec:orbital_variability} for details.}
\label{fig:phase_resolved_qu_obs}
\end{figure}

The X-ray PA shows small but significant variations around the average value; the fit with a constant to the observed PA values gives unacceptable fit with $\chi^2$/dof=33/12, 
suggesting that the spread of PA is higher than statistical noise.
On the other hand, when considering the sample as a whole, there is no systematic trend of PA with hardness -- soft- and hard-sate average PAs are consistent within the uncertainties ($-25\degr\pm1\degr$ and $-24\degr\pm1\degr$, respectively).
The variations of PA might be related to one of the periods of the system: either orbital ($P_{\rm orb}=5.6$~d) or super-orbital ($P_{\rm so}{\sim}300$~d).
Below we investigate variations of PA at these timescales in more detail.

\subsection{Orbital variability}
\label{sec:orbital_variability}

The orbital motion of an X-ray source in the presence of a companion star and circumstellar (intrabinary) matter can induce variations of the observed polarization on the orbital timescale  \citep{BME,Kravtsov2020,Rankin2024,Ahlberg24}. 
To verify the presence of variability at the orbital period in IXPE observations of \cyg, we performed a phase-resolved polarimetric analysis.
The dependence of PD on the spectral hardness prevents us from combining the data in all spectral states.
To minimize the influence of spectral hardness on polarization, we treated soft- and hard-state data separately.

We primarily focus on the hard-state data owing to repeating coverage of the same orbital phases over years and one complete orbital cycle (Epoch 1).
We split the Epoch 1 data into 10 phase bins as described in Sect.\ref{sec:ixpe_reductions}, while the other, shorter observations were not subdivided.
The results are shown in Figs.~\ref{fig:phase_resolved_obs} and \ref{fig:phase_resolved_qu_obs}, where variations of the Stokes $q$ and $u$ are presented as function of the orbital phase and on the $(q,u)$-plane, and also through variation in PD and PA. 
The X-ray polarization of \cyg shows hints of orbital variability: the PD varies sinusoidally, gradually increasing from $\approx$3\% at orbital phase 0.1 to the maximum of $\approx$5\% at phase 0.5. 
The observed PA displays a swing with an amplitude of $\lesssim$10\degr, roughly half-cycle out of phase with the PD. 
This behavior can be interpreted as the tip of the polarization vector tracing a closed loop in the $(q,u)$-plane (see Fig.~\ref{fig:phase_resolved_obs}, left panel). 

\subsubsection{Constant versus sine-wave}

To assess the statistical significance of the orbital variations, we used the F-test to compare two nested models. 
The null hypothesis, $H_0$, assumes that the Stokes parameters $q$ and $u$ remain constant across all orbital phases. 
It has two free parameters: $q_{\rm c}$ and $u_{\rm c}$.
We find the $\chi_0^2$/d.o.f. = 30.5/28 for the fit with this model (see dashed black line in Fig.~\ref{fig:phase_resolved_qu_obs}).

This was tested against the alternative hypothesis, $H_{\rm A}$, in which the Stokes parameters follow simple sinusoidal functions of the orbital phase. 
We notice that $q$ and $u$ have nearly the same amplitude and are out of phase. 
To keep the number of parameters to minimum, we assume the same amplitude $r_{\rm v}$ of the sine wave and we fix the phase shift to $\pi/2$. 
This model can be expressed as follows:
\begin{equation}
\begin{aligned}
    q(\varphi) & = q_{\rm c} + r_{\rm v}\cos(\varphi-\varphi_0),\\
    u(\varphi) & = u_{\rm c} + r_{\rm v}\sin(\varphi-\varphi_0),
\end{aligned}
\end{equation}
where indices (c) and (v) denote constant and variable components, respectively, and $\varphi_0$ is the zero-phase angle.
The model has four free parameters: $q_{\rm c}$, $u_{\rm c}$, $r_{\rm v}$,  and $\varphi_0$.
In this model, the  trajectory the points track at the $(q,u)$-plane may be decomposed into the sum of two vectors: the first one, constant vector, links the origin with the average polarization (center of the circle), and another, varying vector, with constant length of $r_{\rm v}$ and with the azimuthal angle being related to the orbital phase $(\varphi-\varphi_0)$, links the center of the circle with the instantaneous $q,u$ parameters. 
Fig.~\ref{fig:phase_resolved_qu_obs} shows the best-fit model with the solid blue lines that gives $\chi_{\rm A}^2$/d.o.f.=24.2/26.

The resulting value of $F$-statistic, $F = [(\chi_0^2 - \chi_{\rm A}^2)/(\rm{dof}_0 - \rm{dof}_{\rm A})]/[\chi_{\rm A}^2/\rm{dof}_{\rm A}]=3.4$, corresponds to the $p$-value, $p = 1 - \text{CDF} (F) = 0.048$, indicating $4.8$\% probability that the null hypothesis $H_0$ is correct. 
Hence, using this approach we find that orbital variations in the X-ray polarization of \cyg are marginally significant.

\subsubsection{Rotating vector model}

The rather low statistical significance in favor of the sinusoidal model may be related to our assumption that the amplitude of the variable component $r_{\rm v}$ remains constant throughout the orbit. 
However, a number of physical reasons may lead to its variations with orbital phase and over time, leading to the enhanced spread of data points around the average $q,u$ values. 
For example, IXPE has detected well ordered variations of the PA with the spin phase in a number of X-ray pulsars, while the PD showed rather irregular behavior \citep[see][for a recent review]{Poutanen2024}.  
In \cyg too, the dependence of $r_{\rm v}$ on the orbital phase may be complex and include random fluctuations leading to the loss of signal. 

% Fig. 7
\begin{figure}
\centering
\includegraphics[width=\linewidth]{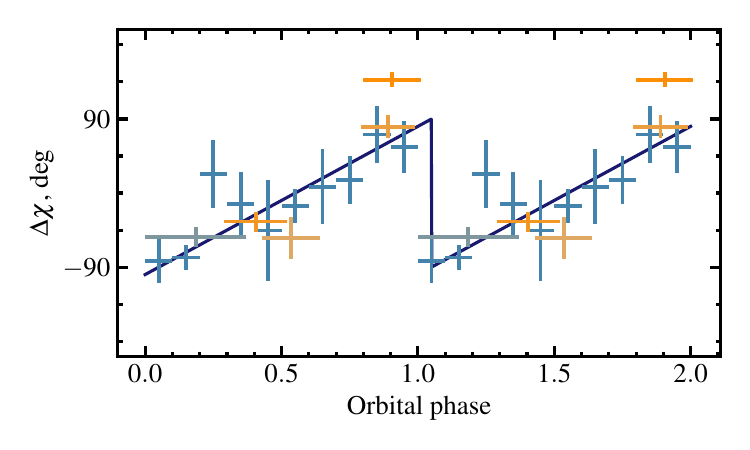}
\caption{Angle $\Delta\chi$ as a function of  the orbital phase in \cyg. 
The solid line shows the best fit to the Epoch 1 data with the linear model \eqref{eq:psilin} described in Sect.~\ref{sec:orbital_variability}. 
}
\label{fig:phase_resolved_around_avg}
\end{figure}

To account for this uncertainty, we adopted an alternative approach to test for the presence of orbital variability by leveraging the expectation that such variations should follow a coherent pattern, specifically, a closed loop in the $(q,u)$-plane, rather than appearing as random statistical fluctuations scattered around the mean.
Below we consider only the orbital phase-dependent changes of PA of the variable component, which does not encompass the aforementioned uncertainty on its PD.
This is similar to the application of the rotating vector model to X-ray pulsar polarimetric data \citep[e.g.,][]{Suleimanov2023,Poutanen2024,Poutanen2024b,Forsblom2025} or searching for rotation of the X-ray PA with time in blazars \citep{DiGesu2023,Kim2024,Pacciani2025}.

From the observed normalized Stokes parameters we subtracted the average values to obtain a new set of Stokes parameters representing the variable component: 
\begin{equation}
q_{\rm v} = q - \langle q \rangle , \quad  u_{\rm v}=u - \langle u \rangle. 
\end{equation} 
We calculated $\langle q \rangle = 3.0 \pm0.1\%$ and $\langle u \rangle = -2.7 \pm0.1\%$ from the Epoch~1 data only (ten blue data points in Fig.~\ref{fig:phase_resolved_obs}), because it continuously covers the whole orbital period -- this ensures that the center point of the loop will not be biased due to unequal orbital coverage. 
We then computed PD$_{\rm v}$ and PA$_{\rm v}$ of the variable component using standard formulae 

\begin{equation}
\mathrm{PD} = \sqrt{q^2 + u^2},\qquad 
\mathrm{PA} = \frac{1}{2}\atantwo(u, q).  
\label{eq:pd_pa}
\end{equation}

The phase-resolved variations of PA$_{\rm v}$ are shown in Fig.~\ref{fig:phase_resolved_around_avg}. 
We see that PA$_{\rm v}$ shows pronounced, gradual $180\degr$ rotation along the orbit (corresponding to one full loop in the left panel of Fig.~\ref{fig:phase_resolved_obs}).
Here, for the sake of an easier interpretation of the observed results, we show $\Delta\chi$, which is the difference between PA$_{\rm v}$ and the  $\langle \rm{PA} \rangle = -21\degr$ of the constant component (obtained from  $\langle q \rangle$ and $\langle u \rangle$ using Eq.~\ref{eq:pd_pa})

\begin{equation}
\Delta\chi = \rm{PA}_{\rm{v}} - \langle \rm{PA} \rangle. 
\end{equation}
One can think of $\Delta\chi$ as the  deviations of the polarization orientation of the variable component from the average direction of polarization on the sky (i.e., relative to the position angle of the orbital axis, rather than relative to North).

When PD$_{\rm v}$ is comparable to its error, the PA$_{\rm v}$ is not normally distributed. 
To compare models to the data, we should use the probability density function of the PA$_{\rm v}$, $\psi$,  from \citet{Naghizadeh1993}:
\begin{equation} \label{eq:PA_dist}
G(\psi) = \frac{1}{\sqrt{\pi}} 
\left\{  \frac{1}{\sqrt{\pi}}  + 
\eta {\rm e}^{\eta^2} 
\left[ 1 + {\rm erf}(\eta) \right]
\right\} {\rm e}^{-p_0^2/2} , 
\end{equation}
where $p_0= p/\sigma_{\rm p}$ is the measured PD$_{\rm v}$ in units of its error (which is just the error on $q$ or $u$ for a given point),  $\eta=p_0 \cos[2(\psi-\psi_0)]/\sqrt{2}$, $\psi_0$ is the measured value for the PA$_{\rm v}$, and \mbox{erf} is the error function. 
The best fit can be obtained by minimizing the log-likelihood function 
\begin{equation}
\label{eq:logL}
    \log L= -2 \sum_{i} \ln   G(\chi_{i}) , 
\end{equation} 
with the sum taken over all phase bins $i$. 

The model to be tested is a linear dependence of the PA with orbital phase: 
\begin{equation}\label{eq:psilin}
\psi_{\rm L} (\varphi)= \psi_0 + \frac{1}{2}\varphi,      
\end{equation}
where $1/2$ comes from the fact  that a full loop of 360\degr\ in the $(q,u)$-plane corresponds to a 180\degr\ change in PA.
If PA$_{\rm v}$ is distributed randomly, then the linear model is not expected to perform better than a constant model  
\begin{equation}\label{eq:psiconst}
\psi_{\rm c}  (\varphi) = \psi_0 .      
\end{equation} 
The best-fit with the linear model gives $\log L=32.7$, while the constant model gives $\log L=37.5$. 
The difference  $\Delta\log L=4.8$ corresponds to the significance of $\exp(-\frac{1}{2}\Delta\log L)=0.09$ that the linear model is preferred.  
As an additional test, we performed Monte-Carlo simulations distributing randomly 15 PAs in the interval $[-90\degr,90\degr]$ and performing a fit with the linear model. 
In 5.2\% cases we got the value of $\log L$ better than what we obtained with the real data for linear model.

The low significance of the linear trend is influenced by the outlier (Epoch 11), and may result from our assumption that the center of the loop in the $(q,u)$-plane remains constant over several years of observation.
Hence, we performed the same analysis only for Epoch 1, which covers one whole orbital cycle.
For Epoch 1, we get $\log L=11.2$ for the linear model, while the constant model gives $\log L=29.9$. 
The difference of $\Delta\log L=18.7$ implies that the orbital variations in Epoch 1 are  significant at the confidence level of $8.7\times10^{-5}$ (${\approx}4\sigma$). 
Monte-Carlo simulations for 10 randomly distributed PAs give 0.2\% chance of getting a value of $\log L$ lower than what we got for real data with the linear model, corresponding to ${\approx}3\sigma$ significance. 
The best-fit to Epoch~1 data with the linear model is shown in Fig.~\ref{fig:phase_resolved_around_avg} with the solid line. 
We note that the data are consistent with only one cycle per orbit, and that the PA rotates counterclockwise.
This contrasts with the behavior of optical polarization in this source, which exhibits two loops per cycle in the $(q, u)$-plane and a clockwise rotation \citep{Kravtsov2023}.

\subsection{Superorbital variability}
 
Next, we tested presence of super-orbital variability in the X-ray polarization data.
The period of these modulations was reported to vary over time: both $\sim$300~d and $\sim$150~d were found \citep[e.g.,][]{Brocksopp1999a,Lachowicz2006, Ibragimov2007, Zdziarski2011}. 
The nature of these variations remains uncertain, but the scenario with precession of the accretion disk or its inner parts is consistent with variations of several observables simultaneously \citep{Bochkarev1983,Poutanen2008}.
If true, this model predicts coordinated variations of the X-ray polarimetric signatures.
These variations, however, can be mixed with the orbital variability, hence polarization from different superorbital phases should be compared for the same orbital bins.
Moreover, because of the change of the average PD between hard and soft states, polarization changes cannot be considered jointly for both states.

To test the precessing inner flow scenario, we folded the hard-state X-ray polarization measurements according to the phase of the superorbital period $P_{\rm so}=294$~d with zero point JD 2440000 \citep{Priedhorsky1983,Kemp1983}.\footnote{No significant variations of X-ray polarization were previously found over ${\sim}73$~d \citep{Krawczynski2022}, which corresponds to the strongest long-term period observed in MAXI \citep{Matsuoka2009} light curves, but had an instrumental origin.}
In Fig.~\ref{fig:phase_resolved_obs} we show the resulting PD and PA color-coded according to the superorbital phase.
Current data show no obvious dependence of the polarization properties on the superorbital phase: for the same orbital phase bins, both PD and PA are consistent with being constant for different superorbital phases.
The only exception is the PA of Epoch 11 (orange cross centered at orbital phase 0.9), which, however, has similar super-orbital phase to the Epochs 9 and 10 (centered at orbital phases 0.9 and 0.4, respectively), which are well aligned with other points at different superorbital phases (blue crosses).
This indicates that the PA of the outlier Epoch 11 is not directly related to the superorbital changes.
Current data suggest that, if present, the superorbital changes of PA do not exceed $\pm 5\degr$ with $3\sigma$ confidence.
This value is substantially smaller than the previously suggested misalignment angle between the orbital and inner flow axes $\sim$15\degr--20\degr\  needed to explain the observed variations of fluxes and polarization signatures.

\begin{figure}
\centering
\includegraphics[width=1\linewidth]{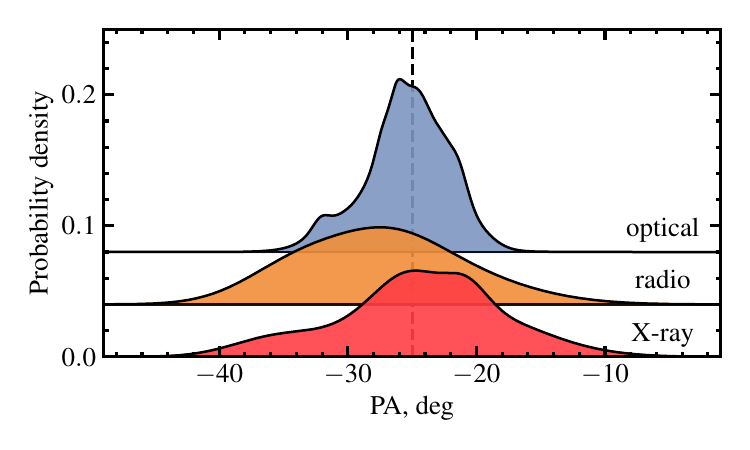}
\caption{Smoothed histogram of the observed PA of \cyg in X-rays (red), radio (orange), and optical (blue). Each observation here is  represented as a Gaussian with the standard deviation equal to the error and the sum of those is divided by the number of observations.
Graphs are shifted vertically for clarity. 
The vertical dashed line shows the radio jet direction.}   \label{fig:pizza_plot_PA_multiwavelength}
\end{figure}

\begin{figure*}
\centering
\includegraphics[width=0.75\linewidth, trim={0cm 8cm 0cm 0cm}]{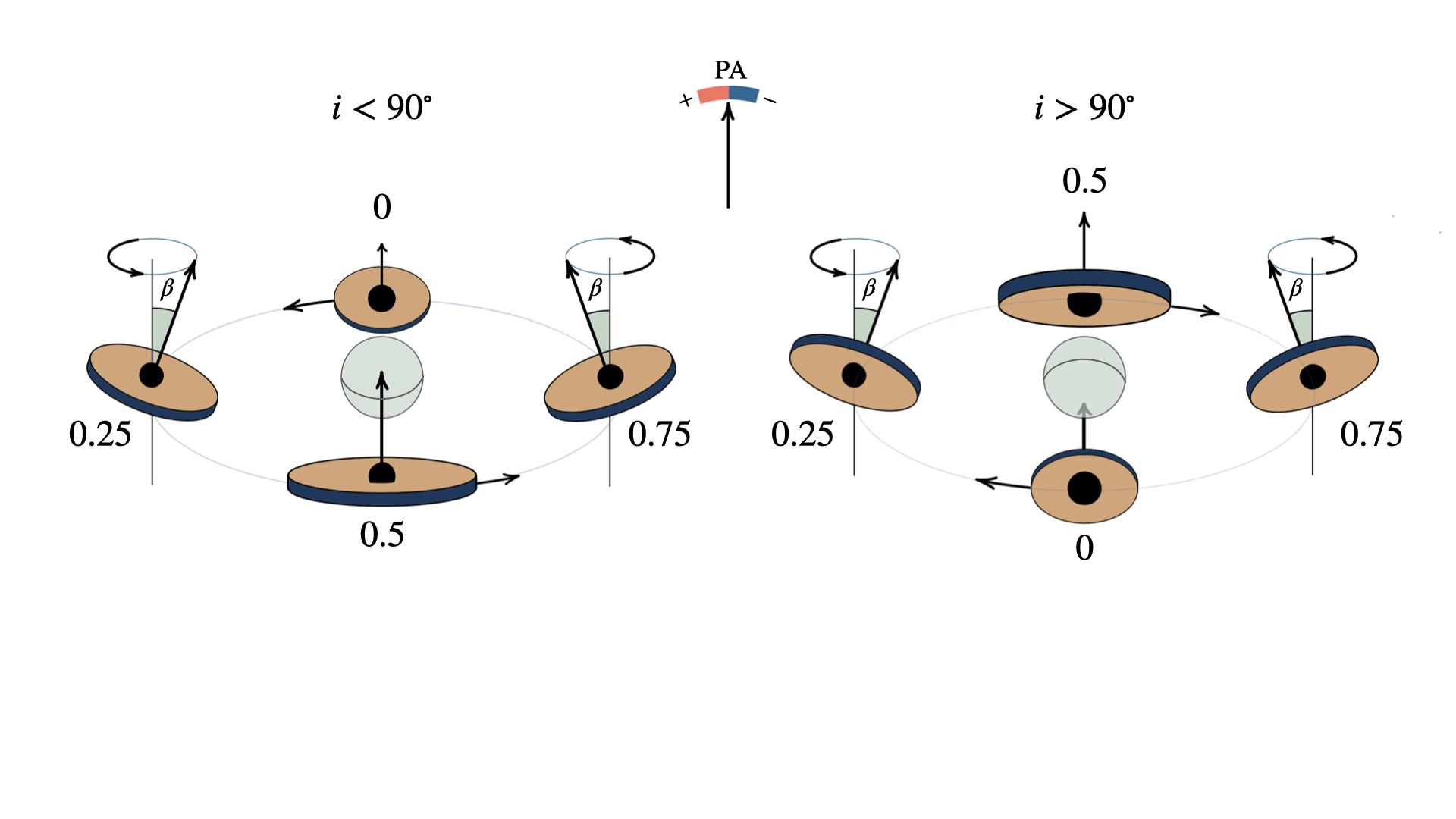}
\caption{Sketch of considered precessing disk geometry for two cases: $i < 90\degr$ and $i > 90\degr$.}
\label{fig:sketch}
\end{figure*}

\subsection{Multiwavelength behavior}

IXPE observations were supported by optical and radio polarimetric observations. 
In Fig.~\ref{fig:multiwavelength} we show the results of multiwavelength polarization evolution.
The average PAs of intrinsic (corrected for interstellar contribution) optical and radio polarization are well aligned with the PA in X-rays (see Fig.~\ref{fig:pizza_plot_PA_multiwavelength}.
At the same time, the average PAs in radio ($\rm{PA}_{\rm R} = -28\degr\pm4\degr$), optical ($\rm{PA}_{\rm O} = -25\degr\pm5\degr$), and X-rays ($\rm{PA}_{\rm X} = -25\degr\pm5\degr$) are all aligned with the jet direction \citep[$\approx$$-25\degr$;][]{Stirling2001,Miller-Jones2021}.

Optical PD and PA are known to show orbital variability, leading to the spread of points visible in Fig.~\ref{fig:multiwavelength}b.
Superorbital variability is known to affect the orbit-average PD and can also be expected to be seen through the shifts of orbit-average PA \citep{Kemp1983,Kravtsov2023}.
Overall, optical polarization does not show any pronounced dependence on the spectral state.

Current radio polarization data are not sufficient to trace any signs of orbital or superorbital variability; however, the PAs of all individual datasets are well aligned with the optical and X-ray PAs.
The radio PD, on the other hand, seems to be sensitive to the state transitions: nearly five-fold increase in PD is observed during the transition to the soft state (around Epoch 13).
At the same time, the X-ray polarization shows a pronounced drop.
Hence, if any connection exists, the radio polarization appears to be anti-correlated with the X-ray polarization.

\section{Discussion}
\label{sec:discussion}

\subsection{Source of orbital variability}
\label{sec:orb_variations}

Several processes may induce orbital variations of the observed X-ray polarization in BHXRBs: reflection of X-rays off the companion star \citep{Ahlberg24, Rankin2024} or from the bow shock (V. Ahlberg et al., in prep.), scattering in the stellar or accretion disk wind \citep{Kallman2015, Nitindala2025}, or changes in the accretion disk orientation \citep{Bochkarev1983}. 
Below we consider two major classes of models: (i) variations of orientation of the region where the X-ray polarization is produced and (ii) scattering of the incident emission at the intrabinary structure. 
The models are constrained by the observed criteria imposed by polarization properties: (a) one loop in the $(q,u)$-plane  per orbital cycle and (b) counter-clockwise rotation (Sect.~\ref{sec:orbital_variability} and the increasing angle with the orbital phase in Fig.~\ref{fig:phase_resolved_around_avg}).

The variations of X-ray polarization can be produced by the changing orientation of the site of X-ray production, we can apply the formalism developed for the rotating vector model \citep{Radhakrishnan1969,Meszaros1988} to constrain the tilt of the X-ray-emitting inner flow (disk) with respect to the orbital plane.
Variations in the PA in this scenario are attributed to changes in the position angle of the projected inner flow axis on the sky. 
If the disk axis changes its orientation as a function of orbital phase, the PA will vary accordingly. 
The limited range of PA variations (Fig.~\ref{fig:phase_resolved_obs}) corresponds to a small tilt, $\beta \approx 2\degr$, which can account for the observed changes.
The observed PD variations are caused by the changing inclination.
We consider a simplified model in which the PD depends on the cosine $\mu$ of the angle between the disk axis and the observer line of sight, given by $\rm{PD} = \rm{PD}_{\rm{max}} (1 - \mu)$.
While the actual angular dependence may be more complex, it can be approximated as a linear function of $\mu$ within a narrow angular range.
The solid black lines in Fig.~\ref{fig:phase_resolved_obs} (loop in the $(q,u)$-plane and sine waves in PD, PA representation) correspond to the best-fit predictions of this model with the tilt $\beta = 2\degr$, $\rm{PD}_{\rm{max}} = 0.37$\% and fixed orbital inclination $i = 153\degr$. 

Changes of the inner disk or flow orientation with respect to the average direction of the orbital axis can be related to tidal forces or precession.
In the case of tidal locking, the orientation of the disk is expected to be fixed with respect to the line connecting the compact object and companion star (e.g., always facing the companion star). 
In this case, the movement of the disk axis on the sky follows the sense of rotation of the compact object around the supergiant (see Fig.~\ref{fig:sketch}).
This, however, is not observed: the criterion (b) implies that rotation of polarization is counter-clockwise, while the binary system is undergoing clockwise rotation on the sky \citep[$i>90\degr$;][]{Miller-Jones2021}.

Alternatively, the disk may experience precession in the direction opposite to the direction of orbital motion (retrograde precession).
Furthermore, if the precession is synchronous with the orbital period, it should proceed in the binary system with companions of comparable masses, low eccentricity and no known third-body influence.
If all these criteria are satisfied, the precession can explain the observed orbital changes of X-ray polarization; however, no known binary system currently fits all of the aforementioned criteria.
Hence, we consider this scenario implausible.

In the second class of models (ii), the changing polarization is produced by the scattering (or reflection) on the intrabinary structure.
The PD in this case depends on the cosine of the scattering angle, $\mu$, as $\rm{PD} = (1 - \mu^2)/(1 + \mu^2)$ (single-scattering case, \citealt{Cha60}).
This relation predicts two PD peaks per orbital period, as $\mu^2$ is identical for all diametrically opposite points along the orbit.
To produce only one peak per orbit and comply with criterion (a), the scattering matter should either be asymmetrically distributed relative to the orbital plane, or the fraction of scattered radiation should vary significantly with the orbital phase.
The latter case has been considered in the context of an asymmetric bow shock producing pronounced orbital variability of X-ray polarization in Cyg~X-3 \citep[][and V. Ahlberg et al. in prep.]{Veledina2024}.
Polarization in this case can be orthogonal to the orbital axis, hence, it is effectively subtracted from the average polarization of the source, leading to higher net PD at phases when the contribution of the scattering is low.
This makes this scenario potentially compliant with criterion (b); however, further quantitative study is needed to draw conclusions on the applicability of this scenario.

\subsection{Long-term trend in polarization}

The long-term stability of the X-ray polarization in \cyg is remarkable and indicates the stability of geometry of the site of polarization production.
The statistically significant increase in PD with energy found in both hard and soft spectral states, along with the constant PA across the IXPE energy band (Sect.~\ref{sec:long-term_behavior}) imposes strong constraints on the Faraday rotation effects within the medium responsible for the production of the observed light and any screen between this site and the observer.
These constraints translate to upper limits on the large-scale magnetic fields, for instance, a large-scale vertical field $B_{z}=10^6$~G can induce the rotation of PA by $\sim$$5\degr$ across the IXPE band \citep{Barnier2024}.
Our updated estimates on the energy dependence of PA can further refine the constraints on the magnitude of $B$-fields in \cyg.

The statistically significant increase in PD with energy detected in the hard state challenges current models of polarization production.
The spectrum is shaped by multiple Compton up-scatterings, with the PD increasing with each successive scattering order \citep[e.g.,][]{Poutanen1996}.
One way to account for the high observed PD (given a fixed system inclination) in this state is to assume that the IXPE band is dominated by high-order Compton scatterings. 
This is equivalent to assuming a low energy for the seed photons, such as those originating from synchrotron radiation \citep[e.g.,][]{Poutanen2009,Malzac2009,Veledina2013}.
However, this scenario leads to a suppression of any energy dependence of polarization, since the PD is known to saturate at high ($\gtrsim$5th) scattering orders \citep{Poutanen1996,Poutanen2023}.
An alternative solution is to retain seed photons from the underlying accretion disc (the slab corona geometry), thus keeping the scattering orders in the IXPE band low to preserve the energy dependence of polarization, and enhance the PD by invoking a bulk matter outflow \citep{Beloborodov1999,Poutanen2023}.
However, realistic outflow velocities have been found insufficient to reproduce the observed $\sim4$\% PD at an inclination $i\approx150\degr$ for this geometry \citep{Poutanen2023}.

The soft-state polarization data are likewise challenging to interpret.
The soft-state PA remains aligned with the jet axis throughout years, in contrast to the early expectations for the polarization produced by the optically thick atmosphere of the disk \citep{Cha60,Sobolev1963,Rees1975,Sunyaev1985}.
Furthermore, the IXPE band detects the Wien part of the disk spectrum (emission beyond the peak of its emission), where the effects of strong gravity and fast matter motion are most prominent and lead to a rotation of PA with energy and depolarization effects that increase with energy \citep[e.g.,][]{ConnorsStark1977,StarkConnors1977,Dovciak2008,Li2009,Loktev2022,Loktev2024}.
We find no evidence for either PA rotation or depolarization; in contrast, the PD is found to increase with energy at a high significance.

Possible scenarios that can reproduce the observed spectra and polarization signatures include a dominant role of returning radiation in the IXPE band \citep{Steiner2024}.
This scenario, however, had been considered in the approximation of the energy-independent reflection albedo of unity, indicating a fully ionized accretion disk near the black hole.
However, the presence of atomic lines is not aligned with this assumption.
Furthermore, the predicted polarization remains below the detected values for the orbital inclination $i\approx153\degr$.

In an alternative scenario, the soft spectrum seen in the IXPE band is composed of both disk emission and the low-temperature Compton scattering continuum.
Such a spectral decomposition implies the presence of both low-energy Maxwellian electrons and high-energy power-law-like electrons (the so-called hybrid Comptonization), and had been previously shown to fit the observed broadband spectrum of \cyg up to MeV energies \citep{Gierlinski1999,Poutanen2009}.
In this case, the expected PD and its increase with energy can be made consistent with the data (A. Bocharova et al. in prep.).

Interestingly, the increasing PD with energy is expected in both scenarios. 
The disk emission generally has lower polarization as compared to the reflected (self-irradiated) part that becomes dominant at energies above the disk peak \citep[][]{Schnittman2009}, leading to the natural increase in PD with energy.
In the hybrid Comptonization scenario, the increase in PD is attributed to both the low disk intrinsic polarization and the presence of the first Compton scattering order that is polarized in the direction along the disk plane \citep{Poutanen1996}.
The dominant role of the disk emission in the IXPE band during the soft state can likewise explain the observed increase of total PD with hardness (Fig.~\ref{fig:pol_vs_hardness}).

\begin{figure}
\centering
\includegraphics[width=0.9\linewidth]{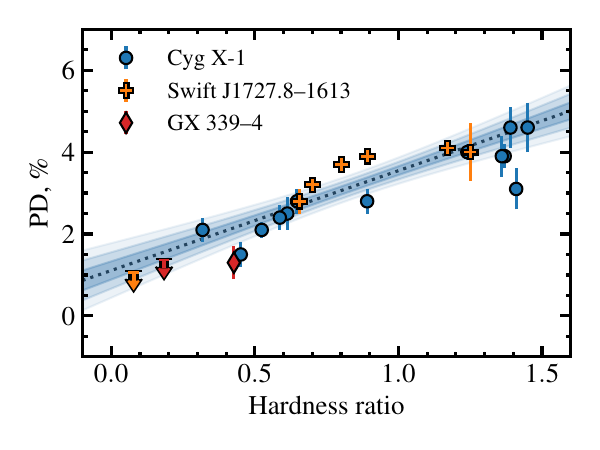}
\caption{Dependence of PD on spectral hardness for \cyg, \swiftJ and \gx. For consistency, the hardness ratio has been calculated the same way for all objects as a ratio of energy flux in 4--8 keV to that in 2--4 keV as measured by IXPE. The dotted line with 1, 2, and 3$\sigma$ confidence intervals shows the linear fit to the data.}   \label{fig:pd_vs_hardness_all_sources}
\end{figure}

In Fig.~\ref{fig:pd_vs_hardness_all_sources} we compare the dependence of the X-ray PD on spectral hardness for sources that are believed to have low or intermediate inclinations: \cyg, \swiftJ and \gx (\citealt{Svoboda2024,Podgorny2024,Mastroserio2025}, see a similar comparison in \citealt{Brigitte2025}).
We note that in all these systems the X-ray PAs are found to be aligned with the jet direction \citep{Krawczynski2022,Ingram2024,Mastroserio2025}.
To enable comparison between systems of various brightness, we computed the spectral hardness the same way for all considered objects as the ratio of the energy flux in 4--8 keV band to that in the 2--4 keV band, as measured by IXPE. 
The polarimetric data have been adopted from \citet{Podgorny2024} and \citet{Mastroserio2025}.
Note that this definition of spectral hardness differs from the hardness ratio used earlier in this work.
The new data on \cyg reveal a PD–hardness dependence similar to that observed in the other BHXRBs, both in slope and in absolute values.
This suggests a remarkable consistency in polarization behavior across different sources and luminosities, for state transitions occurring in both the upper and lower branches of the q-diagram \citep{Belloni2010}. 

\subsection{Multiwavelength view}

The optical polarization is believed to be produced by the scattering of primary star's radiation by the matter surrounding the black hole (i.e. the accretion disk). 
In that case, the average optical PA indicates the orientation of the symmetry axis of the large-scale disk in the sky.
The observed variations of the optical PA with the amplitude of $\Delta \rm{PA} \approx 5\degr$ are caused by the orbital motion of the black hole in the binary \cite[more details can be found in][]{Kravtsov2023}.

Radio polarization in X-ray binaries arises from synchrotron radiation, the dominant emission mechanism within their jets \citep{Westfold1959, Bjornsson1982, Han1992, Corbel2000}. In these systems -- where jets are typically optically thin or partially self-absorbed -- the polarization angle is expected to be orthogonal to the magnetic field responsible for particle acceleration \citep[see, e.g.,][for a review]{HighEnergyAstrophysics}. Although relativistic aberration and internal Faraday effects can complicate this picture \citep[e.g.,][]{Lyutikov2005}, the observed alignment of the radio polarization angle with the jet axis implies a magnetic field structure that is predominantly toroidal, with its projected component on the sky lying perpendicular to the jet. The temporal stability of this polarization angle further suggests a long-lived, ordered magnetic field configuration in the jet’s emission region. A similar alignment in the hard state jets was observed in the BHXRBs \mbox{GRS~1915+105} \citep{Hannikainen2000} and \mbox{MAXI~J1836$-$194}  \citep{Russell2015}, suggestive of a similar magnetic field geometry. 

Following the source transition from hard to soft state (around 2024 December 12; MJD 60656), we observed a fivefold increase in radio polarization degree (PD), from $<$1\% to ${\sim}$5\%. 
During such transitions, BHXRBs are known to disrupt their (steady) hard state jets and launch discrete plasma knots, which -- when polarimetric data are available -- often appear more highly polarized \citep[e.g.,][]{Han1992, Brocksopp2007, Curran2014}. 
Whether this increase is driven by changes in absorption conditions or intrinsic jet structure remains unclear. A spatially resolved ejection in \cyg\ observed by \citet{Fender2006} supports the idea that the PD increase may correspond to such an event. 
A more detailed analysis of radio polarization of \cyg will be given in the follow-up paper (A. K. Hughes et al., in prep.).

Because X-ray polarization likely originates in the innermost regions of the accretion disk, reflecting the geometry of the inner disk and corona, the observed alignment of PAs across radio, optical, and X-ray bands places valuable constraints on the large-scale geometry of \cyg.

\section{Summary}\label{sec:conclusions}

We present the results of a comprehensive three-year observational campaign of \cyg using X-ray, optical, and radio polarimetry.
Thirteen IXPE observations conducted between 2022 and 2024 reveal a clear dependence of the X-ray polarization on the spectral state.
We find the PD to be approximately twice as high in the hard state (${\approx}4.0\%$) as compared to the soft state (${\approx}2.2\%$), and stable over time.

In both states, the PD increases with energy.
The X-ray PA remains independent of the spectral state and shows no clear dependence on photon energy.
Our improved constraints further limit the strength of Faraday effects, providing tight bounds on the magnetic field strength in the region where the X-rays are produced.

We investigate the presence of orbital variability in the X-ray polarization data and find it to be significant for the Epoch~1, which covers the entire orbital period.
Interestingly, the X-ray polarization changes are cyclic with orbital period, corresponding to one loop in the $(q,u)$-plane, with PA rotating counter-clockwise.
This contrasts with the optical polarization behavior, which shows two loops per orbital period and clockwise rotation.
We discuss the potential sources of the X-ray polarization variability and suggest that it may result from scattering of X-ray emission by the circumstellar or intrabinary medium.
Future high-precision X-ray polarimetric observations are essential for further identification of this variability.

We find no evidence of superorbital variability in the X-ray polarization data at the previously reported period of $P_{\rm so}=294$~d.
We place an upper limit of $5\degr$ on the tilt between the orbital axis and the instantaneous axis of the accretion disk or inner flow.
This constraint rules out the earlier hypothesis that the high PD observed in the hard state could be attributed to a favorable phase of superorbital precession, specifically one that would have increased the inclination of the hot inner flow by $15\degr-20\degr$ relative to the orbital axis during the 2022 observation.

We compare \cyg\ with other low- to intermediate-inclination BHXRBs in the PD–spectral hardness diagram and find a remarkable consistency in polarization behavior across different sources and luminosities, on both the upper and lower branches of the q-diagram.
Our multiwavelength data suggest a possible anti-correlation between the X-ray and radio PDs, while no clear correlation is observed with optical polarization.
We show that the average PAs are well aligned across all bands.
These results have broad implications for theoretical models of multiwavelength polarization production in BHXRBs.

\begin{acknowledgements}
The Imaging X-ray Polarimetry Explorer (IXPE) is a joint US and Italian mission. The US contribution is supported by the National Aeronautics and Space Administration (NASA) and led and managed by its Marshall Space Flight Center (MSFC), with industry partner Ball Aerospace (contract NNM15AA18C).  The Italian contribution is supported by the Italian Space Agency (Agenzia Spaziale Italiana, ASI) through contract ASI-OHBI-2022-13-I.0, agreements ASI-INAF-2022-19-HH.0 and ASI-INFN-2017.13-H0, and its Space Science Data Center (SSDC) with agreements ASI-INAF-2022-14-HH.0 and ASI-INFN 2021-43-HH.0, and by the Istituto Nazionale di Astrofisica (INAF) and the Istituto Nazionale di Fisica Nucleare (INFN) in Italy.  This research used data products provided by the IXPE Team (MSFC, SSDC, INAF, and INFN) and distributed with additional software tools by the High-Energy Astrophysics Science Archive Research Center (HEASARC), at NASA Goddard Space Flight Center (GSFC). 
This research has made use of data from the RoboPol programme, a collaboration between Caltech, the University of Crete, IA-FORTH, IUCAA, the MPIfR, and the Nicolaus Copernicus University, which was conducted at Skinakas Observatory in Crete, Greece.
We thank the staff of Lord's Bridge, Cambridge, for their support in making the AMI observations. 

This research has been supported by the Finnish Cultural Foundation (VK), the Academy of Finland grant 355672 (AV) and the Magnus Ehrnrooth foundation (APN). 
MD, JPod, and JS acknowledge the support from the Czech Science Foundation project GACR 21–06825X and the institutional support from the Astronomical Institute RVO:6798581. 
FM, ADM, and FLM are partially supported by MAECI with grant CN24GR08 ``GRBAXP: Guangxi-Rome Bilateral Agreement for X-ray Polarimetry in Astrophysics''.
AI acknowledges support from the Royal Society. 
IL was funded by the European Union ERC-2022-STG-BOOTES-101076343. Views and opinions expressed are however those of the author(s) only and do not necessarily reflect those of the European Union or the European Research Council Executive Agency. Neither the European Union nor the granting authority can be held responsible for them.
SAT is supported by the Ministry of Science and Higher Education of the Russian Federation grant 075-15-2022-262 (13.MNPMU.21.0003). 
MB acknowledges the support from GAUK project No. 102323.
GM acknowledges financial support from the European Union’s Horizon Europe research and innovation programme under the Marie Sk\l{}odowska-Curie grant agreement No. 101107057.
POP acknowledges financial support from the Action Thématique PEM from CNRS and from the french spacial agency CNES.
AAZ has been supported by the Polish National Science Center grants 2019/35/B/ST9/03944 and 2023/48/Q/ST9/00138.

\end{acknowledgements}

\bibliographystyle{aa.bst}
\bibliography{aanda.bib}

\end{document}